\documentclass[a4paper,11pt]{article}
\usepackage{pos}
\usepackage{graphicx}
\usepackage{subcaption}

\newcommand{\ya}{y_{1}}
\newcommand{\yb}{y_{2}}
\newcommand{\za}{z_{1}}
\newcommand{\zb}{z_{2}}
\newcommand{\omya}{\big(1-y_{1}\big)}
\newcommand{\omyb}{\big(1-y_{2}\big)}
\newcommand{\omza}{\big(1-z_{1}\big)}
\newcommand{\omzb}{\big(1-z_{2}\big)}
\newcommand{\omyaomyb}{\big(1-y_{1}\big(1-y_{2}\big)\big)}

\title{Recent progress in antenna subtraction at NNLO and N$^3$LO}

\author*[a]{Matteo Marcoli}

\affiliation[a]{Institute for Particle Physics Phenomenology,\\
	Department of Physics, University of Durham, Durham, DH1 3LE, UK}

\emailAdd{matteo.marcoli@durham.ac.uk}

\abstract{
	In this contribution I will review recent developments in the antenna subtraction method for higher-order calculations in QCD. In particular, I will illustrate the definition and applications of generalised antenna functions for final-state radiation at NNLO and present the first N$^3$LO differential calculation performed entirely with antenna subtraction for jet production at electron-positron colliders. Finally, I will discuss how the extension of generalised antenna functions at N$^3$LO will allow to tackle more complicated processes at this perturbative order.
}

\FullConference{17th International Symposium on Radiative Corrections: Applications of Quantum Field Theory to Phenomenology (RADCOR2025)\\
	5--10 October 2025\\
	Puri, India\\}

\begin{document}
	\maketitle
	
	\section{Introduction}
	
	Precise theoretical predictions for collider observables are essential for exploiting the full potential
	of current and future experiments in high-energy particle collisions. In perturbative QCD, this requires calculations at increasingly high orders in the strong coupling constant, where infrared singularities from soft and collinear radiation pose a major challenge. Slicing techniques and infrared subtraction methods provide systematic frameworks to isolate and cancel these singularities between real and virtual contributions, to achieve physical predictions.
	
	The state-of-the-art of perturbative QCD at NNLO is represented by calculations for so-called \textit{high-multiplicity} processes, namely processes with $2\to3$ kinematics, enabled by the recent evaluation of five-point two-loop amplitudes. Since 2020, NNLO-accurate predictions for phenomenologically relevant $2\to3$ processes were obtained with the sector-improved residue subtraction technique~\cite{Chawdhry:2019bji,Hartanto:2022qhh,Czakon:2021mjy,Alvarez:2023fhi,Chawdhry:2021hkp,Badger:2023mgf,Badger:2025ilt}, the $q_T$-slicing formalism~\cite{Garbarino:2025bfg,Devoto:2024nhl,Buonocore:2023ljm,Buonocore:2022pqq,Catani:2022mfv,Biello:2024pgo,Mazzitelli:2024ura,Kallweit:2020gcp} and the antenna subtraction method~\cite{Chen:2022ktf,Buccioni:2025bkl}. In parallel, significant progress has been made in the development and refinement of infrared subtraction frameworks, with efforts focussing on generalising and automating available methods~\cite{Chen:2022ktf,Gehrmann:2023dxm,Fox:2024bfp,Devoto:2025kin,Devoto:2025eyc,Devoto:2025jql,Bertolotti:2025clg,Bertolotti:2022aih,VanThurenhout:2024hmd,DelDuca:2025yph}, delivering public software for NNLO calculations~\cite{Czakon:2023hls,NNLOJET:2025rno,DelDuca:2024ovc,Devoto:2025cuf}, and exploring promising new avenues~\cite{LTD:2024yrb,Capatti:2025khs,Kermanschah:2021wbk}.
	N$^3$LO calculations pose an even greater challenge. Nevertheless, a series of (semi-)inclusive~\cite{Baglio:2022wzu,Chen:2022vzo,He:2025hin,Duhr:2020sdp,Duhr:2020seh,Duhr:2019kwi,Anastasiou:2015vya,Dreyer:2018qbw,Dreyer:2016oyx} and fully-differential~\cite{Cieri:2018oms,Mondini:2019gid,Currie:2018fgr,Neumann:2022lft,Campbell:2023lcy,Chen:2025kez,Chen:2022xnd,Chen:2022lwc,Chen:2022cgv,Chen:2021vtu,Chen:2021isd} results have already been delivered. The latter are among the most challenging computations in perturbative QCD, and are available only for simple processes: colour-singlet production or decay and DIS. The techniques capable to reach such high perturbative accuracy can not be straightforwardly extended beyond the class of processes mentioned above. New approaches are then required to tackle N$^3$LO calculations for arbitrary processes, in particular in the presence of final-state jets. 
	
	Antenna subtraction~\cite{Gehrmann-DeRidder:2005btv,Currie:2013vh} has proven to be a powerful and flexible method for NNLO calculations, in particular for processes involving jets. The method exploits the universal factorisation properties of QCD in unresolved limits and organises the singular behaviour of real-emission matrix elements into so-called \textit{antenna functions}~\cite{Gehrmann-DeRidder:2004ttg,Gehrmann-DeRidder:2005alt,Gehrmann-DeRidder:2005svg}, which represent the building blocks of the subtraction formalism. In this contribution, we summarise two recent advances in the antenna subtraction formalism. First, we discuss the construction of \textit{generalised antenna functions} at NNLO for final-state radiation~\cite{Fox:2024bfp}, which enable an automated construction of NNLO infrared counterterms and a more efficient numerical evaluation. Second, we review the application of antenna subtraction to jet production in electron-positron annihilation at N$^3$LO~\cite{Chen:2025kez,Chen:2025ojp}.
	
	\section{Generalised antenna functions for final-state radiation at NNLO}
	
	Standard antenna functions~\cite{Gehrmann-DeRidder:2004ttg,Gehrmann-DeRidder:2005alt,Gehrmann-DeRidder:2005svg} describe the singular behaviour of real-emission matrix elements
	in configurations where one unresolved parton is emitted between two hard radiators.
	They are traditionally extracted from colour-ordered matrix elements for simple processes, or,
	more recently, constructed via the \textit{designer antenna algorithm}~\cite{Braun-White:2023zwd,Braun-White:2023sgd}, which yields simpler expressions with better isolation of infrared limits and the removal of unphysical singularities.
	
\begin{figure}[t]
		\centering
		\begin{subfigure}[t]{0.22\textwidth}
			\centering
			\includegraphics[width=\textwidth]{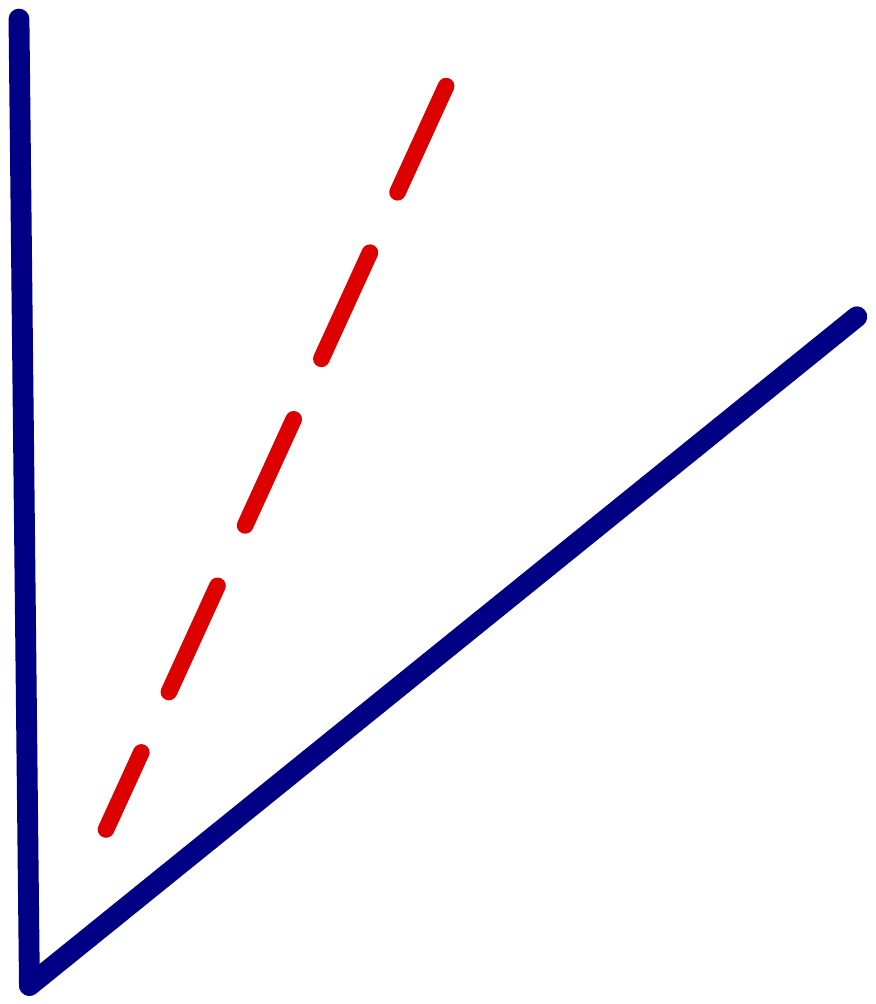}
			\caption{}
			\label{fig:topo_NLO}
		\end{subfigure}
		\hfill
		\begin{subfigure}[t]{0.22\textwidth}
			\centering
			\includegraphics[width=\textwidth]{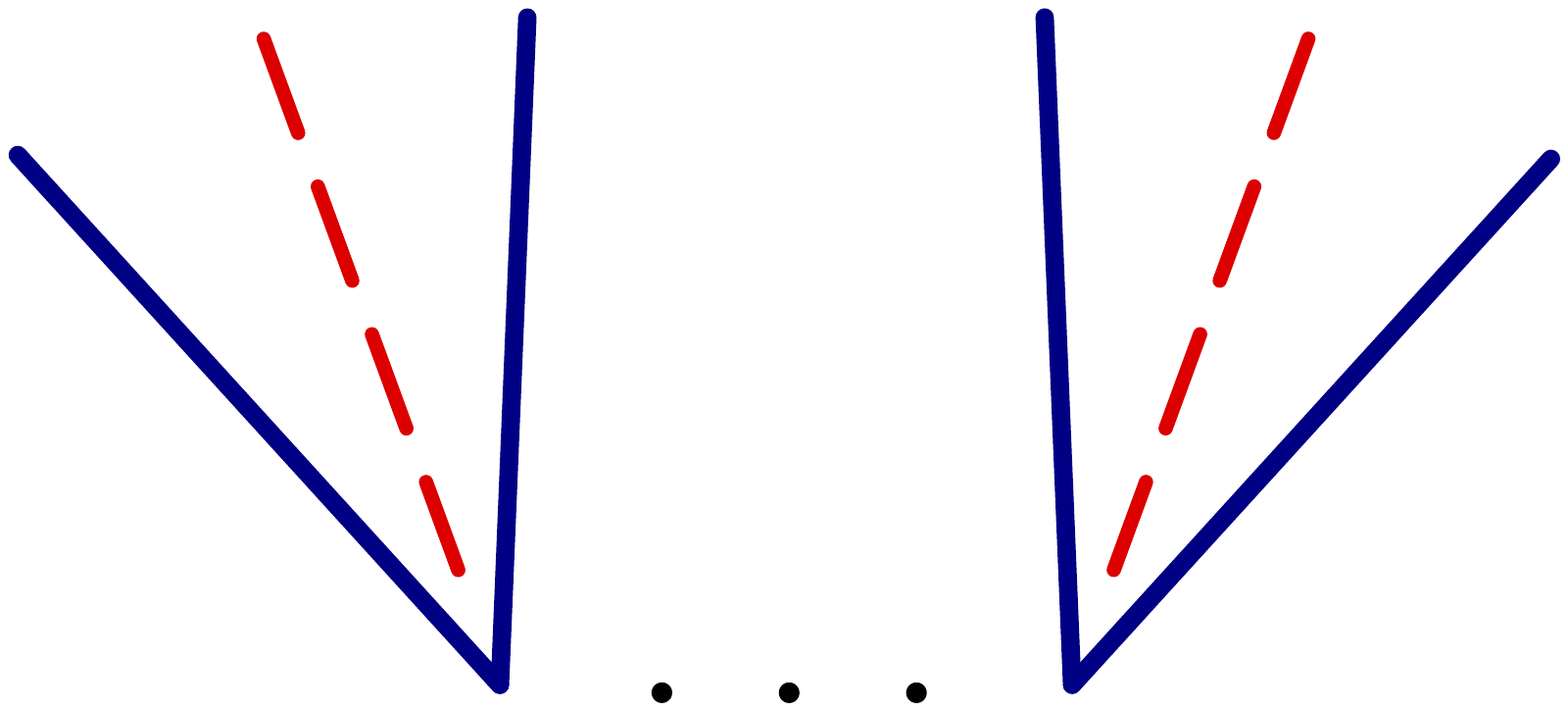}
			\caption{}
			\label{fig:topo_unconnected}
		\end{subfigure}
		\hfill
		\begin{subfigure}[t]{0.22\textwidth}
			\centering
			\includegraphics[width=\textwidth]{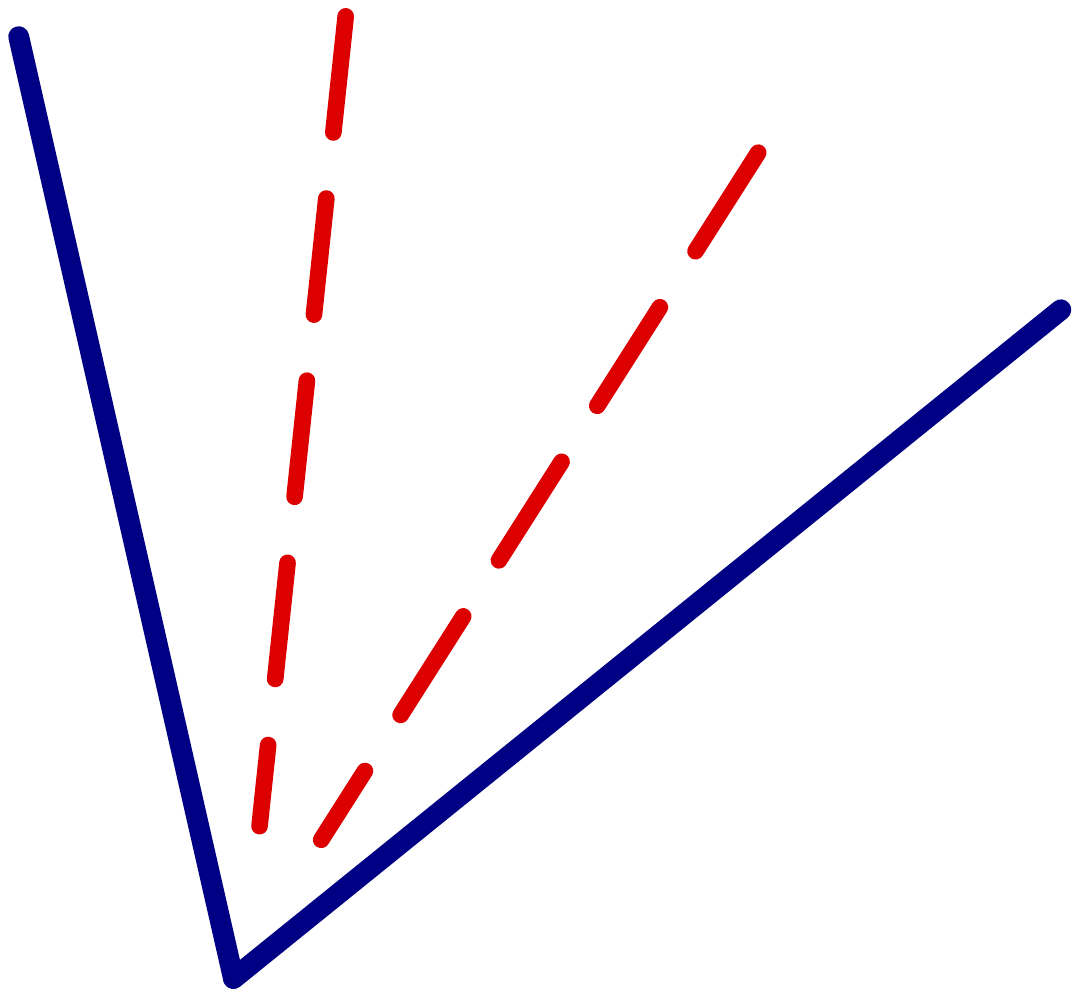}
			\caption{}
			\label{fig:topo_connected}
		\end{subfigure}
		\hfill
		\begin{subfigure}[t]{0.22\textwidth}
			\centering
			\includegraphics[width=\textwidth]{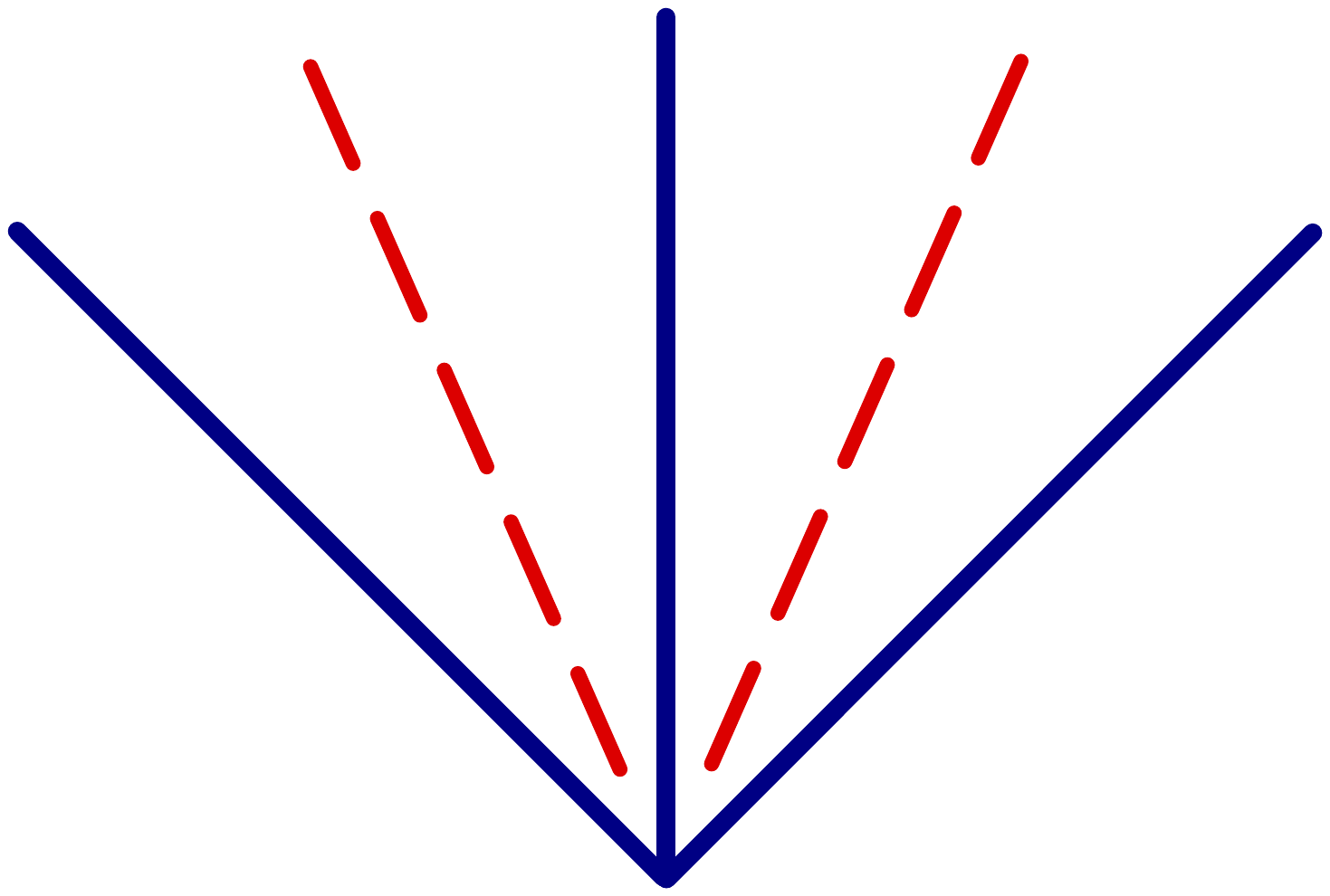}
			\caption{}
			\label{fig:topo_almost}
		\end{subfigure}
		\caption{Emission topologies at NLO (a) and NNLO (b)--(d). Blue lines denote hard radiators; red dashed lines denote unresolved emissions. The almost colour-connected topology (d) is the most problematic and motivates the introduction of generalised antenna functions.}
		\label{fig:topologies}
	\end{figure}

	At NLO, a single unresolved emission always involves exactly one pair of hard radiators, so standard
	antenna functions provide a complete and efficient description. At NNLO, however, two unresolved emissions
	can arise in three qualitatively distinct topologies, illustrated in Fig.~\ref{fig:topologies}.
	In the \textit{colour-unconnected} case [panel (b)], the two emissions share no hard radiator and the
	subtraction is handled by a fully iterated product of two standard NLO antennae. In the
	\textit{colour-connected} case [panel (c)], both hard radiators are shared and the configuration is
	captured by a standard four-parton antenna function, supplemented by iterated corrections to avoid
	double-counting of single-unresolved behaviour~\cite{Gehrmann-DeRidder:2005btv,Currie:2013vh}. Both of these topologies are well handled
	within the existing formalism.
	The most problematic topology is the \textit{almost colour-connected} one [panel (d)], where only one
	hard radiator is shared between the two emissions. This configuration is not fully iterated: the two
	unresolved partons feel each other through the recoil on the common radiator. Traditional antenna
	functions can be applied, but this requires a very complicated sequence of iterated structures
	supplemented by large-angle soft terms~\cite{Weinzierl:2008iv,Currie:2013vh}, leading to the most complex and numerically
	inefficient sector of antenna subtraction at NNLO.
	To address this, generalised antenna functions involving \textit{five} partons and \textit{three} hard radiators were introduced
	in~\cite{Fox:2024bfp}. Using the designer antenna algorithm, it is possible to construct antenna functions
	$X^0_{5,3}(i^h, j, k^h, l, m^h)$ that describe two unresolved emissions between three hard radiators (labelled with the $h$ subscript)
	via a single compact object. 
	The $X^0_{5,3}$ antenna functions require a $5\to3$ momentum mapping: $(i^h, j, k^h, l, m^h)\to(I^h,K^h,M^h)$. We choose to employ an iterated dipole mapping of the form:
	\begin{alignat}{2}
		\label{eq:X53Mmapping}
		&p_I  =&   p_i+p_j - \frac{s_{ij}}{s_{ik}+s_{jk}} p_k, \nonumber \\
		\text{map}_{5\to 3}: \hspace{1cm}    &p_K  =&  \left(1+ \frac{s_{ij}}{s_{ik}+s_{jk}} +\frac{s_{lm}}{s_{lk}+s_{mk}} \right)p_k, \\ 
		&p_M  =& p_l+p_m - \frac{s_{lm}}{s_{lk}+s_{mk}} p_k, \nonumber ,
	\end{alignat}
	where $s_{ij}=2p_i\cdot p_j$.
	A key feature of this mapping is that it factorises the phase space in a way that makes analytical
	integration particularly tractable. Specifically, the integration of the $X^0_{5,3}$ antenna functions over the phase space of the two unresolved emissions reduces to a single parametric integral of the form
	\begin{align}\label{eq:integral}
		&I(b_1,b_2,b_3,b_4,b_5,b_6,b_7,b_8,b_9) = 
		\int_0^1 
		\za^{b_1} \omza^{b_2} d\za\,
		\int_0^1 \zb^{b_3} \omzb^{b_4} 
		d\zb\, \nonumber \\
		& \hspace{4cm}\times
		\int_0^1\int_0^1 \ya^{b_5} \omya^{b_6} 
		\yb^{b_7} \omyb^{b_8} \omyaomyb^{b_9} d\ya\, d\yb\,
		\nonumber \\
		&\qquad=
		\frac{\Gamma(1+b_1) \Gamma(1+b_2)}{\Gamma(2+b_1+b_2)}
		\frac{\Gamma(1+b_3) \Gamma(1+b_4)}{\Gamma(2+b_3+b_4)}
		\frac{\Gamma(1+b_5) \Gamma(1+b_6)}{\Gamma(2+b_5+b_6)}
		\frac{\Gamma(1+b_7) \Gamma(1+b_8)}{\Gamma(2+b_7+b_8)}
		\nonumber \\
		&\hspace{4cm} \times     {}_{3}F_{2} ([1+b_5,1+b_8,-b_9],[2+b_5+b_6,2+b_7+b_8],1).
	\end{align}
	With this result, a fully-analytical subtraction scheme at NNLO whcih employs the novel antenna functions can be developed~\cite{Fox:2024bfp}.
	Generalised antenna functions have been validated at NNLO against the original formulation for event
	shapes in $e^+e^-$ annihilation, finding perfect numerical agreement while achieving a subtraction that
	is significantly simpler in structure and up to 5--10 times faster numerically~\cite{Fox:2024bfp}. 
	%This is illustrated in Fig.~\ref{fig:validation}, which shows the ratio of the NNLO correction to event shape observables computed with the new and the original method at each individual colour factor.
	Applications to hadronic Higgs decays, including jet rates at order $\alpha_s^3$ (three-jet rate at
	NNLO and two-jet rate at N$^3$LO) and event shapes at NNLO, have also been presented~\cite{Fox:2025qmp,Fox:2025cuz}. Recently, the generalised antenna functions were employed to perform the first NNLO calculation of four-jet production at electron-positron colliders~\cite{Chen:2025kez}.

%	\begin{figure}[t]
%		\centering
%		\includegraphics[width=0.4\textwidth]{fig_validation.pdf}
%		\caption{Ratio of NNLO event-shape coefficients for $e^+e^-\to jjj$ computed with the new generalised
%			antenna function approach over the original method, shown for several observables and at each
%			independent colour factor. The ratio is consistent with unity throughout, demonstrating the
%			equivalence of the two approaches. Figure from Ref.~\cite{Fox:2024bfp}.}
%		\label{fig:validation}
%	\end{figure}
	
	\section{Jet production in electron-positron annihilation at N$^3$LO with antenna subtraction}
	
	The process $e^+e^- \to jj$ at N$^3$LO represents the first fully-differential fixed-order calculation
	for jet production at this perturbative order performed entirely with a local infrared subtraction
	method~\cite{Chen:2025kez,Chen:2025ojp}. The calculation builds on the existing $e^+e^- \to jjj$ NNLO implementation in
	\textsc{NNLOJET}~\cite{NNLOJET:2025rno} and employs standard matrix-element-based antenna functions, whose analytical
	integration over the final-state phase space at N$^3$LO was carried out in~\cite{Jakubcik:2022zdi}.
	The process is particularly suited as a starting point because it involves only quark-antiquark antenna
	configurations and dipole-like colour correlations at N$^3$LO, limiting the complexity.
	
	At N$^3$LO, the physical cross section is decomposed into four components corresponding to the triple-virtual (VVV), real-double-virtual (RVV), double-real-virtual (RRV), and
	triple-real (RRR) contributions. The subtraction terms needed to remove the infrared singularities of each contribution are built from combinations of integrated and unintegrated N$^3$LO antenna functions, which include five-parton tree-level antenna functions,
	four-parton one-loop antenna functions, and three-parton two-loop antenna functions~\cite{Chen:2025kez}. A major technical challenge
	is the numerical stability of the one-loop double-unresolved quantities appearing in the RRV sector
	and the two-loop single-unresolved quantities in the RVV sector. These are handled by rescue systems
	that trigger quadruple-precision evaluation and, respectively, Taylor expansions of the relevant
	special functions in numerically unstable phase-space regions.
	
\begin{figure}[t]
		\centering
		\includegraphics[width=0.4\textwidth]{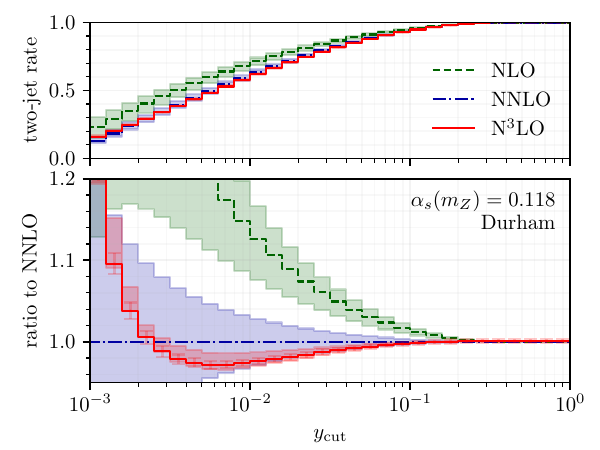}
		\caption{Two-jet rate $R_2(y_\text{cut})$ as a function of the Durham jet resolution parameter
			$y_\text{cut}$, computed at LO, NLO, NNLO and N$^3$LO with $\alpha_s(m_Z) = 0.118$. The lower
			panel shows the ratio to the NNLO result. Figure from Ref.~\cite{Chen:2025kez}.}
		\label{fig:R2}
	\end{figure}

	We first computed the N$^3$LO correction to the inclusive hadronic cross
	section, which was found to be $\sigma^{(3)} = \sigma_0 (\alpha_s/2\pi)^3 (-105 \pm 11)$, where $\sigma_0$ represents the Born-level result, in good agreement
	with the exact analytic result of $\sigma^{(3)} = \sigma_0 (\alpha_s/2\pi)^3(-102.14)$~\cite{Chetyrkin:1996ela}. The Durham two-jet rate $R_2^{(3)}(y_\text{cut})$ (Fig.~\ref{fig:R2})
	at order $\alpha_s^3$, computed directly and differentially for the first time, is in full agreement with
	the indirect determination of~\cite{Gehrmann-DeRidder:2008qsl}, which exploits the knowledge of the inclusive cross section at N$^3$LO. These results demonstrate both the correctness of the subtraction
	scheme and the feasibility of fully-differential N$^3$LO calculations with antenna subtraction.
	The almost colour-connected topology discussed in Section~2 --- which requires generalised antenna
	functions with three hard radiators --- only becomes relevant at higher multiplicity, and will be part of the more complicated processes tackled next, such as $e^+e^- \to 3j$ at N$^3$LO. The extension of
	generalised antenna functions to cover N$^3$LO configurations is therefore a crucial next step in the antenna subtraction programme.
	
	\section{Conclusions and outlook}
	
	We have summarised two recent advances in the antenna subtraction formalism. Generalised antenna
	functions with three hard radiators~\cite{Fox:2024bfp} provide a unified, compact, and numerically efficient
	description of almost colour-connected double-unresolved configurations at NNLO, removing the most
	complex sector of the traditional formulation and yielding up to an order-of-magnitude speedup. The
	first application of antenna subtraction to a fully-differential N$^3$LO calculation --- jet
	production in $e^+e^-$ annihilation~\cite{Chen:2025kez} --- demonstrates the viability of antenna subtraction at
	this perturbative order and provides a solid foundation for extending the method to more complex
	processes.
	The natural outlook is the gradual extension of both developments: the generalised antenna functions
	need to be extended to include initial-state radiation and the antenna structures required at N$^3$LO,
	while the N$^3$LO calculation framework will be extended to processes with more final-state partons.
	The combination of generalised antenna functions with the \textit{colourful antenna subtraction method}~\cite{Chen:2022ktf,Gehrmann:2023dxm}
	offers a promising route towards a general, automated infrared subtraction scheme at N$^3$LO.
	
	\section*{Acknowledgements}
	We would like to thank our collaborators Xuan Chen, Elliot Fox, Nigel Glover, Petr Jakub{\v{c}}{\'\i}k and Giovanni Stagnitto. The authors used Claude (Anthropic) as a writing assistant for drafting portions of this proceedings contribution. All content was reviewed and approved by the authors.
	
	\bibliographystyle{JHEP}
	\bibliography{radcor}
	
\end{document}